\documentclass[preprint,aps,showpacs,preprintnumbers,amsmath,amssymb,11pt,sort&compress]{revtex4}

\usepackage{fleqn,epsf}
\usepackage{graphicx}
\usepackage{psfrag}
\newcommand{\be}{\begin{equation}}
\newcommand{\ee}{\end{equation}}
\newcommand{\bea}{\begin{eqnarray}}
\newcommand{\eea}{\end{eqnarray}}

\def\lsim{\raise0.3ex\hbox{$\;<$\kern-0.75em\raise-1.1ex\hbox{$\sim\;$}}}
\def\gsim{\raise0.3ex\hbox{$\;>$\kern-0.75em\raise-1.1ex\hbox{$\sim\;$}}}
\begin{document}
\preprint{ }
\title{Two-photon decay of heavy quarkonium \\ from  heavy-quark spin symmetry\footnote{
Talk given at the 
{\em QCD@Work 2007 International Workshop on QCD: Theory and Experiment}, Martina Franca, 
Italy, 16--20 June 2007}}
\author{
T. N. Pham}
\affiliation{
 Centre de Physique Th\'{e}orique, CNRS \\ 
Ecole Polytechnique, 91128 Palaiseau, Cedex, France }
\date{\today}

\begin{abstract}
With the recent measurements on $\eta_{c}$ and $\eta_{c}^{\prime}$
at CLEO, Babar and Belle, and with the prospect of finding the $\eta_{b}$
at the Tevatron, it seems appropriate to have another look at the two-photon
decay of heavy quarkonium from the standpoint of an effective Lagrangian
based on local operator expansion and heavy-quark spin symmetry.
In this talk, I would like to discuss a recent work on the two-photon
decay rates of ground states and excited states of $\eta_c$ and
$\eta_b$ using the local operator expansion approach and heavy-quark spin
symmetry and taking into account the binding-energy. We find
that the predicted two-photon width for $\eta_c$ agrees well with experiment, 
but the predicted value for  $\eta_c(2S)$ is twice larger than the CLEO 
estimation. We point out that the essentially model-independent ratio 
of $\eta_b$ two-photon width to the $\Upsilon$ leptonic width and 
the $\eta_b $ two-photon 
width could be used to extract the strong coupling constant $\alpha_s$ . 
\end{abstract}


\maketitle

\section{Introduction}
The  physics of quarkonium decay seems to be rather well 
understood within the conventional framework of QCD
\cite{Brambilla,Colangelo}. 
 However a recent experimental estimation of
the two-photon width of the $\eta'_c$ 
by the CLEO collaboration \cite{Asner} gives 
$\Gamma_{\gamma \gamma}(\eta'_c)= 1.3
\pm 0.6$ keV,  which contradicts most of the existing theoretical 
predictions  in the range $3.7$ to $5.7$
keV~\cite{Ackleh,Kim,Ahmady,Chao,Munz,Ebert,Gerasimov,Crater}.
This is surprising, since the non-relativistic 
$\eta_{c}^{\prime}\to \gamma \gamma$ decay rate differs from that
for $\eta_{c}$ only by the wave function at the origin:  thus it is
difficult to lower the predicted value without considering other
effects like  binding effects and relativistic corrections. 
In fact since the  first excited  state  $\eta^\prime_c$ is more that 
$600\,\rm MeV$ above the $\eta_c$, the mass effects on the decay rate 
could be important and a better approach would be to use relativistic 
kinematics in the calculation
of the decay rate. In this talk, I would like to discuss a recent 
work\cite{Lansberg} in which we use heavy-quark spin
symmetry(HQSS) and an effective Lagrangian from  local operator 
product expansion to 
relate the two-photon width of  charmonium and bottomonium singlet states 
to  the leptonic width of the triplet  states. We find that, the
predicted two-photon width of $\eta_c$ agrees with experiment, but  
the predicted value for $\eta'_c$ is twice the CLEO value. For $\eta_{b}$,
$\eta_{b}^{\prime}$ and $\eta_{b}^{''}$, the predicted widths 
are higher than most of the existing calculations. 

\section{Effective Lagrangian for $^{1}S_{0}$ Decay into two-photon}

In the two-photon decay of a heavy quarkonium bound state, the 
outgoing-photon momentum is large compared with the relative momentum 
of the quark-antiquark bound state, which is $O(\Lambda/m_{Q}) $, with
$\Lambda \ll m_{Q}$. One obtains an effective Lagrangian for the process 
$c\bar{c}\to \gamma\gamma$ (represented by the upper diagram in 
Fig.\ref{Fig:decay}) by expanding the heavy quark propagator,  
{\em e.g.} charm quark, in powers of $q^{2}/m_{c}^{2}$, and  neglecting 
${\cal O}(q^{2}/m_{c}^{2})$ terms ($q=p_{c}-p_{\bar{c}}$). Like leptonic
decay, the two-photon decay, in this approximation, is described  
by the following effective Lagrangian:
\begin{figure}[hbp]
\centering
\leavevmode
\epsfxsize=8cm
\epsffile{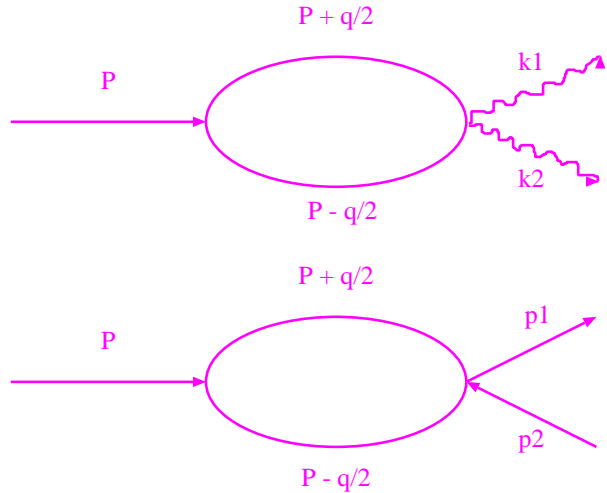}
\caption{effective coupling between a $c\bar{c}$ and two-photon (upper)
  and a lepton pair }
\label{Fig:decay}
\end{figure}
\bea
&&{\cal L}^{\gamma \gamma}_{\rm eff}=-i c_1(\bar{c}\gamma_{\sigma}\gamma_{5} c) \varepsilon_{\mu \nu \rho \sigma} F_{\mu\nu}
A_\rho \nonumber \\
&&{\cal L}^{\ell \bar \ell}_{\rm eff}=-  c_2(\bar{c} \gamma_{\mu} c) (\ell \gamma_\mu \bar \ell)
\label{Leff}
\eea
with 
\be
c_1\simeq \frac{Q_c^2 (4\pi
  \alpha_{\rm em})}{M_{\eta_c}^2+b_{\eta_c} M_{\eta_c}} , \quad \quad 
c_2=\frac{Q_c (4\pi \alpha_{\rm em})}{M_\psi^2} \,\,\, .
\label{coeff}
\ee
The factor $1/(M_{\eta_c}^2+b_{\eta_c} M_{\eta_c})$ in $c_{1}$ 
contains the binding-energy effects (the binding-energy $b$ is defined as
$b =2m_{c} -M$) and is obtained  from the 
denominator of the charm-quark 
propagator ($k_{1}, k_{2}$  being the outgoing-photon momenta):
\be
\frac{1}{[(k_{1}-k_{2})^{2}/4 - m_{c}^{2}]} \,\,\, .
\ee
 The  decay amplitude  is then 
given by the matrix element of the axial vector current
$\bar{c}\gamma_{\mu}\gamma_{5}c$ similar to the quarkonium leptonic
decay amplitude which is given by  the vector current matrix element 
$\bar{c}\gamma_{\mu}c$ for $J/\psi \to e^{+}e^{-}$. Thus for decays of 
 $S$-wave quarkonium into two photons or   a dilepton pair 
$\ell \bar \ell$, we have:
\bea
&&{\cal M}_{\ell \bar \ell}= 
Q_c (4 \pi \alpha_{\rm em})\frac{f_{\psi}}{M_\psi} \varepsilon_\mu (\ell
\gamma^\mu \bar \ell) \nonumber\\
&&{\cal M}_{\gamma \gamma}=- 4 i Q_c^2 (4 \pi \alpha_{\rm em})\frac{f_{\eta_c}}{M^2_{\eta_c}+b_{\eta_c} M_{\eta_c}} \varepsilon_{\mu \nu \rho \sigma} 
\varepsilon_1^\mu \varepsilon_2^\nu k_1^\rho  k_2^\sigma
\label{amp}
\eea
where  
\be
 \left<0|\bar c \gamma_\mu  c| \psi \right>= f_{\psi}M_\psi\varepsilon^\mu, \quad
 \left<0|\bar c \gamma_\mu \gamma_{5} c| \eta_c \right>= i f_{\eta_c}\,P_{\mu}.
\label{dc}
\ee
from which the decay rates are:
\be
\Gamma_{\ell \bar \ell}(\psi)=\frac{4 \pi Q_c^2 \alpha^2_{\rm em}
  f_\psi^2}{3 M_\psi}, \qquad
\Gamma_{\gamma \gamma}(\eta_c)=\frac{4 \pi Q_c^4 \alpha^2_{em} 
f_{\eta_c}^2}{M_{\eta_c}} .
\label{rate}
\ee
   By taking  $M_{\psi}f_\psi^2=12 |\psi(0)|^2$, we recover the usual
non-relativistic expression for the decay rate which, with NLO QCD
 radiative corrections, are given by 
\bea
&&\Gamma^{NLO}(^3S_1)= \Gamma^{LO}(^3S_1) \left(1- \frac{\alpha_s}{\pi}\frac{16}{3}\right) \\ 
&&\Gamma^{NLO}(^1S_0)= \Gamma^{LO}(^1S_0) \left(1- \frac{\alpha_s}{\pi}\frac{(20-\pi^2)}{3}\right)\,\,\, .
\label{rate1}
\eea

\section{Matrix elements of Local operators }

We have shown that in the  approximation of neglecting 
${\cal O}(q^{2}/m_{c}^{2})$ terms, the two-photon
decay amplitude is given by the $\eta_{c}$ decay constant
$f_{\eta_{c}}$. We now  
derive a symmetry relation between $f_{\eta_{c}}$ and  
$f_{J/\psi}$, the  $J/\psi$ leptonic decay constant using the relativistic 
spin projection operators 
for a relativistic Bethe-Salpeter quarkonium bound state.
 
Consider now the matrix elements of local operators in a
fermion-antifermion system with a given spin $S$ and orbital angular 
momentum $L$ \cite{Kuhn,Guberina} :
\be
{\cal A} = \int
\frac{d^{4}\,q}{(2\,\pi)^{4}}\,{\rm Tr}\,{\cal O}(0)\chi(P,q)
\label{A1}
\ee
$P$ is the total 4-momentum of the quarkonium system, $q$ is the relative
4-momentum between the quark and anti-quark and $\chi(P,q) $ is
the Bethe-Salpeter wave function.
For a quarkonium system in  a  fixed total, orbital
and spin angular momentum,  $\chi(P,q) $ is given by (${\bf q}$ is the
relative 3-momentum vector of $q$).
\bea
\kern -0.5cm\chi(P,q; J,J_{z},L,S)\kern -0.2cm&& = \sum_{M,S_{z}}\,2\,\pi\delta(q^{0} - \frac{{\bf
    q}^{2}}{2\,m})\psi_{LM}({\bf q})<LM;SS_{z}|JJ_{z}>\nonumber \\
\kern -0.5cm&& \times\sqrt{\frac{3}{m}}\sum_{s,\bar{s}}\,u(P/2 +q,s)\bar{v}(P/2-q,\bar{s})
<\frac{1}{2}s;\frac{1}{2}\bar{s}|S S_{z}> \nonumber \\
\kern -0.5cm&&= \sum_{M,S_{z}}\, 2\,\pi\,\delta(q^{0} - \frac{{\bf
    q}^{2}}{2\,m})\psi_{LM}({\bf q}){\cal P}_{S S_{z}}(P,q)<LM;SS_{z}|JJ_{z}> \,\,\, .
\label{chi}
\eea
The spin projection operators  ${\cal P}_{S S_{z}}(P,q) $ are 
\bea
&&\kern -0.5cm {\cal P}_{0,0}(P,q) = \sqrt{\frac{3}{8m^{3}}}[-(\rlap/P/2 +\rlap/q) +
m]\gamma_{5}\,[(\rlap/P/2 -\rlap/q) + m] \nonumber \\
&&\kern -0.5cm {\cal P}_{1,S_{z}}(P,q) = \sqrt{\frac{3}{8m^{3}}}[-(\rlap/P/2 +\rlap/q) +
m]\rlap/{\varepsilon}(S_{z})\,[(\rlap/P/2 -\rlap/q) + m]  \,\,\, .
\label{PJ}
\eea
  For $S$-wave quarkonium in a spin singlet $S=0$ and spin
triplet $S=1$ state:

\be
{\cal A}(^{2S +1}S_{J}) = {\rm Tr}\,({\cal O}(0)\,{\cal P}_{J\,J_{z}}(P,0))
\int \frac{d^{3}\,q}{(2\,\pi)^{3}}\,\psi_{00}(q) \,\,\, .
\label{A12}
\ee

In this expression  the $q$-dependence in the spin projection operator
has been dropped  and the integral 
in Eq.(\ref{A12}) is the $S$-wave function at the origin \cite{Guberina}:
\be
\int \frac{d^{3}\,q}{(2\,\pi)^{3}}\,\psi_{00}(q) =
\frac{1}{\sqrt{4\,\pi}}{\cal R}_{0}(0) \,\,\,\, .
\label{R0}
\ee

  Using Eq.(\ref{PJ}) and Eq.(\ref{A12}) to compute 
the matrix elements $\left<0|\bar c \gamma_\mu \gamma_{5} c| P \right>$
and $\left<0|\bar c \gamma_\mu  c| V \right>$ for the  singlet $S=0$
pseudo-scalar meson $P$ and for the  triplet $S=1$ vector meson $V$, 
we find, neglecting
quadratic $O(q^{2})$ terms.
\be
f_{P} = \sqrt{\frac{3}{32\,\pi\,m^{3}}}\,{\cal R}_{0}(0)\, (4\,m) \ , \qquad
f_{V} = \sqrt{\frac{3}{32\,\pi\,m^{3}}}\,{\cal R}_{0}(0)\, \frac{(M^{2}
+ 4\,m^{2} )}{M}
\label{fPV}
\ee
   $f_{P}/f_{V}$ (from Eq.(\ref{fPV}) ) is only
quadratic in the binding-energy $b$, and is of the order $O(b^{2}/M^{2})$.

Thus the relation $f_{P}\simeq f_{V}$ is valid to a good approximation.
It is expected that this relation holds also for excited state of charmonium
and bottomonium where the binding terms $O(b^{2}/M^{2})$ can be neglected.
This is a manifestation of heavy-quark spin symmetry(HQSS). In this limit,
the two-photon width of singlet $S=0$ quarkonium state can be obtained
from the leptonic width of triplet $S=0$ quarkonium state 
without using a bound state description.  This approach  differs 
from the traditional non-relativistic bound state approach  in
the use of local operators for which the matrix elements could be
measured or extracted from  physical quantities, or computed
from QCD sum rules \cite{Novikov,Reinders} and lattice QCD \cite{Dudek}.

   The ratio of the $\eta_{c}$ two-photon width
to $J/\psi$ leptonic width in the limit of HQSS is then:
\be
R_{\eta_{c}} = \frac{\Gamma_{\gamma \gamma}(\eta_c)}{\Gamma_{\ell \bar \ell}(J/\psi)}=3\,Q_c^2\,\frac{M_{J/\psi}}{M_{\eta_c}}\left(1+\frac{\alpha_s}{\pi}\frac{(\pi^{2}- 4)}{3}\right)\,\,\, .
\label{Rc}
\ee
For $M_{\eta_c}= M_{J/\psi} $, the above expression becomes the 
usual non-relativistic result \cite{Kwong, Pancheri} as mentioned above.
From  the measured $J/\psi$ leptonic width, we get
$\Gamma_{\gamma \gamma}(\eta_c)=7.46$ keV. Including  NLO QCD 
radiative corrections with $\alpha_{s}=0.26$, we find
$\Gamma_{\gamma \gamma}(\eta_c)=9.66$ keV
in agreement  with the world everage value $7.4 \pm 0.9 \pm 2.1$ keV.
A similar result is obtained in \cite{Pancheri} who gives 
$8.16 \pm 0.57 \pm 0.04$ keV .

Thus the  effective Lagrangian
approach  successfully predicts the $\eta_c$ two-photon width in
a simple, essentially model-independent manner. 

\section{HQSS predictions for $\Gamma_{\gamma \gamma}(\eta^{\prime}_c)$}

 To obtain the prediction for $\eta^{\prime}_c$, we shall apply HQSS
to $2S$ state. Thus, assuming  $f_{\eta^{\prime}_c}=f_{\psi^{\prime}}$, 
and neglecting binding-energy terms,  we find:
$\Gamma^{}_{\gamma \gamma}(\eta'_c)= \Gamma^{}_{\gamma \gamma}(\eta_c) 
\frac{f^2_{\psi'}}{f^2_{J/\psi}}=3.45~\hbox{keV}$. 
This value is  more than twice 
the evaluation by CLEO , but nearly in agreement  
with other theoretical calculations\cite{Ackleh,Kim,Ahmady} as shown in
Table 1. Including binding-energy terms,
for $M_{\eta_{c}}\simeq M_{J/\psi}$, 
$M_{\eta_{c}'}\simeq M_{\psi'}$, we have
\be
\Gamma_{\gamma \gamma}(\eta'_c)
=\Gamma_{\gamma \gamma}(\eta_c)
\left(\frac{1+b_{\eta_c}/ M_{\eta_c}}{1+b_{\eta'_c}/M_{\eta'_c}}\right)^2 
\times \left(\frac{\Gamma_{e^+e^-}(\psi')}{\Gamma_{e^+e^-}(J/\psi)}\right)
\label{be1}
\ee
which gives 
\be
\Gamma_{\gamma \gamma}(\eta'_c) = 4.1\,{\rm keV}\,\,\,\, .
\label{beeffect}
\ee

\begin{table}[h]
\begin{tabular}{cccccccccccc}
\hline
$\Gamma_{\gamma \gamma}$ & This work &\cite{Ackleh} &
\cite{Kim}&\cite{Ahmady} &\cite{Munz}  &\cite{Chao} &\cite{Ebert}\kern -0.3cm&\cite{Crater}\\
\hline
$\eta_c$  & $7.5-10$ & $4.8$ & $7.14 \pm 0.95$&$11.8\pm 0.8\pm 0.6$&$3.5\pm 0.4$&$5.5$&$5.5$ &6.2 \\
$\eta'_c$ & $3.5-4.5$ & $3.7$&$4.44\pm 0.48$&$5.7\pm 0.5\pm 0.6$&
$1.38 \pm 0.3$ &$2.1$&$1.8$&3.36-1.95\\
\hline

\end{tabular}
\caption{  Theoretical predictions for 
$\Gamma_{\gamma \gamma}(\eta_c)$ and $\Gamma_{\gamma \gamma}(\eta'_c)$. (All values
are in units of keV).}\label{tab-res}
\end{table}

  The measured values are from  \cite{pdg} (PDG) and from CLEO\cite{Asner}~:
\bea
&&\Gamma_{\gamma \gamma}(\eta_{c}) = 7.4 \pm 0.9 \pm 2.1 \,{\rm keV},
\quad {\rm PDG}  \nonumber \\
&&\Gamma_{\gamma \gamma}(\eta_{c}') = 1.3 \pm 0.6 \,{\rm keV},
\quad {\rm CLEO}  .
\eea
The CLEO extraction of $ \Gamma_{\gamma \gamma}(\eta_{c}')$ is done by
considering the following quantity :
\be
R(\eta_{c}'/\eta_{c}) = \frac{\Gamma_{\gamma \gamma}(\eta_{c}')\times
  {\cal B}(\eta_{c}' \to K_{S}K\pi)}{\Gamma_{\gamma \gamma}(\eta_{c})\times
  {\cal B}(\eta_{c} \to K_{S}K\pi)} = 0.18\pm0.05\pm 0.02 \,\,\,\, .
\label{CLEO}
\ee

  To obtain  $\Gamma_{\gamma \gamma}(\eta_{c}') $ from
the above data, CLEO assumes 
\be
{\cal B}(\eta_{c}' \to K_{S}K\pi) \approx {\cal B}(\eta_{c} \to K_{S}K\pi)
\label{KKp}
\ee
and finds 
\be
\Gamma_{\gamma \gamma}(\eta_{c}') = 1.3 \pm 0.6 \,{\rm keV} \,\,\,\, .
\label{etac'}
\ee

  On the other hand, Belle measurements of $B\to \eta_{c}K$ and 
$B\to \eta_{c}'K$ gives \cite{Choi}:
\be
R(\eta_{c}'K/\eta_{c}K)= \frac{{\cal B}(B^{0}\to \eta_{c}'K^{0})\times {\cal B}(\eta_{c}'\to
  K_{S}K^{+}\pi^{-})}{{\cal B}(B^{0}\to \eta_{c}K^{0})\times {\cal
    B}(\eta_{c}\to K_{S}K^{+}\pi^{-})} = 0.38\pm 0.12 \pm 0.05
\label{Belle}
\ee

  Using the approximate equality  Eq.(\ref{KKp}), one
would obtain
\be
\frac{{\cal B}(B^{0}\to \eta_{c}'K^{0})}{{\cal B}(B^{0}\to
  \eta_{c}K^{0}) } \approx 0.4
\label{Belle1}
\ee
which agrees more or less with the QCD factorization(QCDF)
prediction \cite{Song}~:
\be
\frac{{\cal B}(B^{0}\to \eta_{c}'K^{0})}{{\cal B}(B^{0}\to
  \eta_{c}K^{0}) } \approx 0.9\times 
(\frac{f_{\eta_{c}'}}{f_{\eta_{c}}})^{2} \approx 0.45 \,\,\, .
\label{qcdf}
\ee

  The  extracted Belle value Eq.( \ref{Belle}) 
is close to the  ratio obtained from the Babar measured
 $B^{+}\to \eta_{c}K^{+} $ and $B^{+}\to \eta_{c}'K^{+} $
branching ratio\cite{pdg} .
\be
\frac{{\cal B}(B^{+}\to \eta_{c}'K^{+})}
{{\cal B}(B^{+}\to  \eta_{c}K^{+})}  = 0.38\pm 0.25
\label{Babar}
\ee

  This is expected since from  $SU(2)$ flavor 
symmetry, one would have  the near equality
between the ratios  ${\cal B}(B^{0}\to \eta_{c}'K^{0})/{\cal B}(B^{0}\to
  \eta_{c}K^{0})  $ and  ${\cal B}(B^{+}\to \eta_{c}'K^{+})/
{\cal B}(B^{+}\to  \eta_{c}K^{+})$. 

  Thus the assumption of
the approximate  equality between the $ \eta'_c \to KK\pi$ and 
$ \eta_c \to KK\pi$ branching ratio seems to be justified to some extent.
This implies the small $\eta'_{c}\to \gamma\gamma$
decay rate quoted above. We note however that the good agreement with 
 QCDF predictions for the measured ratio 
${\cal B}(B^{0}\to \eta_{c}'K^{0})/{\cal B}(B^{0}\to  \eta_{c}K^{0}) $ and 
${\cal B}(B^{+}\to \eta_{c}'K^{+})/{\cal B}(B^{+}\to  \eta_{c}K^{+})$
at Belle and Babar suggests that
$f_{\eta_{c}^{\prime}}/f_{\eta_{c}} \approx f_{\psi'}/f_{J/\psi}$,  
which in turn supports our predicted value for the 
$\eta_{c}^{\prime}$ two-photon width which is  more than twice bigger
than the CLEO estimated value  shown above. More precisely, comparing 
$R(\eta_{c}'/\eta_{c})$  with $R(\eta_{c}'K/\eta_{c}K)$  and using QCDF value
given in Eq.(\ref{qcdf}), we find
\be
R(\eta_{c}'/\eta_{c}) \approx R(\eta_{c}'K/\eta_{c}K)/0.9 \,\,\,\, .
\label{Rt}
\ee
 The Belle data in Eq.(\ref{Belle})  would then implies 
$R(\eta_{c}'/\eta_{c})\approx 0.42\pm 0.13\pm 0.05 $, twice
bigger than the CLEO data shown in Eq.(\ref{CLEO}).
\section{ HQSS predictions for $\Gamma_{\gamma \gamma}(\eta_b)$ and
$\Gamma_{\gamma \gamma}(\eta^{\prime}_b)$ }

  Since the $b$-quark mass is significantly
higher than the  $c$-quark mass, the effective Lagrangian and HQSS 
approach should work better for bottomonia decays  to leptons and photons.
We thus have:
\be
R_{\eta_{b}} = \frac{\Gamma_{\gamma \gamma}(\eta_b)}{\Gamma_{\ell \bar \ell}(\Upsilon)}=3\,Q_b^2\,\frac{M_{\Upsilon}}{M_{\eta_b}}\left(1+\frac{\alpha_s}{\pi}\frac{(\pi^{2}- 4)}{3}\right)
\label{Rb}
\ee
(neglecting the small $b_{\eta_{b}}/M_{\eta_b} $ binding-energy term).
This gives $\Gamma_{\gamma \gamma}(\eta_b)= 560\,\rm eV $ ($\alpha_s(M_{\Upsilon})=0.16$, $M_{\eta_b}=9300$ MeV) .

  For $\eta_{b}'$ and higher excited state, one has 
($M_{\eta_b} \simeq M_{\Upsilon}$ and $M_{\eta_{b}'}\simeq M_{\Upsilon'}$):
\be
\Gamma_{\gamma \gamma}(\eta'_b)
=\Gamma_{\gamma \gamma}(\eta_b)
\left(\frac{1+b_{\eta_b}/ M_{\eta_b}}{1+b_{\eta'_b}/M_{\eta'_b}}\right)^2 
 \left(\frac{\Gamma_{e^+e^-}(\Upsilon')}{\Gamma_{e^+e^-}(\Upsilon)}\right).
\label{etab}
\ee
which gives $\Gamma_{\gamma \gamma}(\eta'_b)= 250\,\rm eV $
and $\Gamma_{\gamma \gamma}(\eta''_b)= 187\,\rm eV $. In Table. \ref{tab-res1}
we give our prediction for the two-photon width of $\eta_{b}$, 
$\eta_{b}^{\prime} $ and $\eta''_{b}$  together with other  
theoretical predictions. We note that our predicted values are 
somewhat higher than these predicted values. 

  Eq.(\ref{Rb}) can be used to determine in a reliable way 
the value of $\alpha_{s}$ . The momentum scale at which  
$\alpha_{s}$ is to be evaluated here  could be in principle be fixed 
with $R_{\eta_b}$. 

  Further check of consistency of the value for 
$\alpha_{s}$ may be possible in future mesurements on the $\eta_{b}$
and its two-photon decay branching ratio: 
\be
 \frac{\Gamma_{\gamma\gamma}(\eta_b)}{\Gamma_{gg}(\eta_{b})}=
\frac{9}{2}\,Q_b^4\,\frac{\alpha^{2}_{em}}{\alpha^{2}_{s}}\left(1- 7.8\,\frac{\alpha_s}{\pi}\right) \,\,\, .
\label{br}
\ee
\begin{table}[h]
\begin{tabular}{cccccccccccc}
\hline
$\Gamma_{\gamma \gamma}$ & This work  &\cite{Schuler} &\cite{Lakhina}&\cite{Ackleh}& \cite{Kim}&
\cite{Ahmady} & \cite{Munz} &\cite{Ebert}&\cite{Godfrey}&\cite{Fabiano}&\cite{Penin} \\
\hline
$\eta_b$ &   $560$  & $460$ & $230$&$170$& $384 \pm 47$& $520$ &$220 \pm 40$& $350$ &$214$&
$466\pm 101$&$659\pm92$\\
$\eta'_b$ &  $269$ & $200$& $70$& - & $191 \pm 25$& - & $110 \pm 20$& 150& 121&-&-\\
$\eta''_b$&  $208$ & -    & $40$ & - & - & -& $84 \pm 12$& 100& 90.6& -&-\\
\hline
\end{tabular}
\caption{ Summary of theoretical predictions for 
$\Gamma_{\gamma \gamma}(\eta_b)$, $\Gamma_{\gamma \gamma}(\eta'_b)$ and 
$\Gamma_{\gamma \gamma}(\eta''_b)$. (All values
are in units of eV).}\label{tab-res1}
\end{table}

\section{Conclusion}
 We have shown in this work that effective Lagrangian approach and HQSS 
can be used to compute quarkonium decays into lepton and photon with 
relativistic kinematic, for both ground state and excited state of 
heavy quarkonium systems. Our predicted $\eta_{c}'\to \gamma\gamma$ 
width is larger than the CLEO estimated value, though 
many relativistic calculations  could give a smaller 
value for $\eta_{c}'\to \gamma\gamma$ width but also produce smaller 
value for the $\eta_{c}\to \gamma\gamma$ width. 

  Measurements of the two-photon widths for $\eta_{b}$
and  higher excited states  could provide a test for HQSS
and a determination of the $\alpha_{s}$ coupling constant at the scale
around the $\Upsilon$ mass, as has been done with the $\Upsilon$ leptonic
width in the past.

\section{Acknowledgments}
I would like to thank G. Nardulli, P. Colangelo, F. De Fazio and
the  organizers  for the warm hospitality extended to me at Martina Franca.

\end{document}